\documentstyle[12pt,sprocl,epsfig]{article}
     
\newcommand{\gtrsim}{~\raisebox{-0.2em}{$\stackrel{\textstyle>}{\sim}~$}}
\newcommand{\lesssim}{~\raisebox{-0.2em}{$\stackrel{\textstyle<}{\sim}~$}}
\begin{document}

%%%%%%%%%%%%%%%%%%%%%%%%

\hfill {\small \bf IFUSP-P1295} 

\hfill {\small\bf IFT-P.009/98}

\vspace{.2in}

\title{Present and Future Searches for Leptoquarks}

%%%\thanks{Talk given                                                          
%%%    by O.\ J.\ P.\ \'Eboli at the International Workshop on ``Physics       
%%%    Beyond the Standard Model: from Theory to Experiment'', Valencia, 1997.}

\author{Oscar J.\ P.\ \'Eboli$^1$, R.\ Z.\ Funchal$^2$, and T.\ L.\ 
  Lungov$^1$  \\[8pt]
  {\em $^1$ Instituto de F\' {\i}sica Te\'orica -- UNESP \\
    R. Pamplona 145, 01405--900 S\~ao Paulo, Brazil \\[8pt]
    $^2$ Instituto de F\'{\i}sica, Universidade de S\~ao Paulo, \\
    C.\ P.\ 66.318, 05315-970 S\~ao Paulo, Brazil. }}

\maketitle

\vspace{.2in}

\hfuzz=25pt

\begin{abstract}
  
  We review the present searches for scalar leptoquarks and the
  potential of the CERN Large Hadron Collider (LHC) to unravel the
  existence of first generation leptoquarks.

\end{abstract}

    Talk given                                                           
    by O.\ J.\ P.\ \'Eboli at the International Workshop on ``Physics       
    Beyond the Standard Model: from Theory to Experiment'', Valencia, 1997.

%******************************************************************************
\section{Introduction}

Many extensions of the Standard Model (SM) treat quarks and leptons in
the same footing and, consequently, allow the existence of particles,
called leptoquarks, that mediate quark-lepton transitions.  The class
of theories exhibiting leptoquarks includes composite \cite{comp,af},
grand unified \cite{gut}, and technicolor \cite{tec} models.

A natural hypothesis for theories beyond the SM is that they exhibit
the gauge symmetry $SU(2)_L \otimes U(1)_Y$ above the electroweak
symmetry breaking scale $v$, therefore, we imposed this symmetry on
the leptoquark interactions.  In order to avoid strong bounds coming
from the proton lifetime experiments, we required baryon ($B$) and
lepton ($L$) number conservation.  The most general effective
Lagrangian for leptoquarks satisfying the above requirements and
electric charge and color conservation is given by \cite{buch}
\begin{eqnarray}
{\cal L}_{{eff}}~  &=& {\cal L}_{F=2} ~+~ {\cal L}_{F=0} 
\; , 
\label{e:int}
\\
{\cal L}_{F=2}~  &=& g_{{1L}}~ \bar{q}^c_L~ i \tau_2~ 
\ell_L ~S_{1L}+ 
g_{{1R}}~ \bar{u}^c_R~ e_R ~ S_{1R} 
+ \tilde{g}_{{1R}}~ \bar{d}^c_R ~ e_R ~ \tilde{S}_1
\nonumber \\
&& +~ g_{3L}~ \bar{q}^c_L~ i \tau_2~\vec{\tau}~ \ell_L \cdot \vec{S}_3 
\; ,
\label{lag:fer}
\nonumber \\
%\label{eff} \\
{\cal L}_{F=0}~  &=& h_{{2L}}~ R_{2L}^T~ \bar{u}_R~ i \tau_2 ~
 \ell_L 
+ h_{{2R}}~ \bar{q}_L  ~ e_R ~  R_{2R} 
+ \tilde{h}_{{2L}}~ \tilde{R}^T_2~ \bar{d}_R~ i \tau_2~ \ell_L
%\; ,
\nonumber
\label{eff} 
\end{eqnarray}
where $F=3B+L$, $q$ ($\ell$) stands for the left-handed quark (lepton)
doublet, and we omitted the flavor indices of the leptoquark couplings
to fermions. The leptoquarks $S_{1R(L)}$ and $\tilde{S}_1$ are
singlets under $SU(2)_L$, while $R_{2R(L)}$ and $\tilde{R}_2$ are
doublets, and $S_3$ is a triplet.

Low-energy experiments lead to strong indirect bounds on the couplings
and masses of leptoquarks.  The main sources of indirect constraints
are:

$\bullet$ Leptoquarks can give rise to Flavor Changing Neutral Current
(FCNC) processes if they couple to more than one family of quarks or
leptons \cite{shanker,fcnc}. In order to avoid strong bounds from
FCNC, we assumed that the leptoquarks couple to a single generation of
quarks and a single one of leptons.

$\bullet$ The analyses of the decays of pseudoscalar mesons put
stringent bounds on leptoquarks unless their coupling is chiral
\cite{shanker}.

$\bullet$ Leptoquarks that couple to the first family of quarks and
leptons are strongly constrained by atomic parity violation
\cite{apv}.  In this case, there is no choice of couplings that avoids
the strong limits.

$\bullet$ The analyses of the effects of leptoquarks on the $Z$
physics through radiative corrections lead to limits on the masses and
couplings of leptoquarks that couple to top quarks \cite{gbjkm}.

As a rule of a thumb, the low-energy data constrain the masses of
leptoquarks to be larger than $0.5$---$1$ TeV when their Yukawa
coupling is equal to the electromagnetic coupling $e$
\cite{gbjkm,leurer}.

%%%%%%%%%%%%%%%%%%%%%%%%%%%%%%%%%%%%%%%%%%%%%%%%%%%%%%%%%%%%%%%%%%%%%%

\section{Present Bounds on Leptoquarks}

Since leptoquarks are an undeniable signal of physics beyond the SM,
there have been several direct searches for them in accelerators.
Recently the LEP Collaborations \cite{lepnew} used their $\sqrt{s}=
161$ and 172 GeV data to obtain the constraint $M_{lq} \gtrsim 131$
GeV for leptoquarks coupling to first family quarks and electrons.
The searches for scalar leptoquarks decaying exclusively into
electron-jet pairs at the Tevatron constrained their masses to be
$M_{lq} \gtrsim 225$ GeV \cite{PP}.

Using the $e^-p$ data, the experiments at HERA \cite{HERA} placed
limits on leptoquark masses and couplings, establishing that $M_{lq}
\gtrsim 216-275$ GeV depending on their types and couplings. These
constraints are stronger for $F=2$ leptoquarks since these can be
formed by $e^-$--valence quark collisions in $e^-p$ interactions. On
the other hand, $e^+p$ collisions at HERA exhibited an intriguing
excess of neutral current events at large $Q^2$ ($\gtrsim 1.5 \times
10^4$ GeV$^2$) \cite{heranew}. H1 observes 18 events for lepton--jet
invariant masses between 187.5 and 212.5 GeV, while only 1.5 events
are expected. ZEUS found 5 events for $x>0.55$ and $y>0.25$ when 2
events are expected. These rates are certainly incompatible with the
standard model.

At this point it is natural to verify whether this excess can be
explained by scalar leptoquarks with the interactions given by Eq.\
\ref{e:int}. For simplicity let us assume that the leptoquarks couple
either to $u$ or $d$ quarks but not to $s$.  From the event rates, the
leptoquark must be $F=0$, otherwise its signal would have already been
observed in the $e^-p$ run. Combining the HERA rates and the atomic
parity violation bounds we obtain that branching ratio to charged
leptons ($B_e$) must satisfy the limit \cite{alta}
\begin{eqnarray}
 B_e \gtrsim \mbox{0.1--0.2} && \mbox{for $e^+u$ leptoquarks} \;\; ,
\label{be1}
\\
 B_e \gtrsim \mbox{0.2--0.4} && \mbox{for $e^+d$ leptoquarks} \;\; .
\label{be2}
\end{eqnarray}

The HERA data suggests that the leptoquark mass is around 200 GeV,
which is within the mass range explored by D\O\ and CDF. The
Tevatron data for leptoquark masses around 200 GeV constrains the 
charged lepton branching ratio ($B_e$) to be
\begin{equation}
B_e \lesssim \mbox{0.5--0.7} \;\; .
\label{be3}
\end{equation}
Therefore the HERA data can be interpreted as being the signal for
scalar leptoquarks provided the $B_e$ lies in the window defined by
Eqs.\ (\ref{be1}), (\ref{be2}), and (\ref{be3}). It is important to
notice that leptoquarks whose interactions are totally described by
Eq.\ (\ref{e:int}) can {\bf not} explain the HERA data since all
leptoquarks that couple to $e^+u$ and $e^+d$ pairs possess $B_e =1$.

%%%%%%%%%%%%%%%%%%%%%%%%%%%%%%%%%%%%%%%%%%%%%%%%%%%%%%%%%%%%%%%%%%%%%%

\section{Future LHC Bounds}
 
In the near future, the direct search for leptoquarks with masses
above a few hundred GeV can be carried out only at the CERN Large
Hadron Collider \cite{fut:pp}.  We studied the capability of the LHC
to unravel the existence of scalar leptoquarks through the final state
topology two jets plus a pair $e^+e^-$ \cite{efl}. This was
accomplished by a careful analyses of the signal and backgrounds using
the event generator PYTHIA \cite{pyt}.  We performed our analyses for
first generation leptoquarks whose interactions are described by the
most general effective Lagrangian that is invariant under $SU(3)_C
\otimes SU(2)_L \otimes U(1)_Y$ given by Eq.\ (\ref{e:int}).

In hadronic colliders, leptoquarks can be single or pair produced
through the processes $q + g \rightarrow S_{{lq}} + \ell$, $q +
\bar{q} \rightarrow S_{{lq}} + \bar{S}_{{lq}}$, and $g + g
\rightarrow S_{{lq}} + \bar{S}_{{lq}}$, where $\ell = e^\pm$ ($\nu$)
and we denoted the scalar leptoquark by $S_{{lq}}$. When the
leptoquark decays into a $e^\pm$--$q$ pair, the final state presents a
pair $e^+e^-$ and jets after hadronization. The cross sections for
pair production are model independent because the leptoquark--gluon
interaction is determined by the $SU(3)_C$ gauge invariance. On the
other hand, the single production is model dependent once it involves
the the unknown Yukawa coupling of leptoquarks to a lepton--quark
pairs.

At the parton level, the single production of leptoquarks leads to a
final state exhibiting a pair $e^+e^-$ and $q$ ($\bar{q}$).  After the
parton shower and hadronization the final state usually contains more
than one jet, and consequently, the backgrounds for single and pair
productions of leptoquarks are basically the same.  In our analyses we
kept track of the $e^\pm$ (jet) carrying the largest transverse
momentum, that we denoted by $e_1$ ($j_1$), and the $e^\pm$ (jet) with
the second largest $p_T$, that we called $e_2$ ($j_2$).

Within the scope of the SM, there are many sources of backgrounds
leading to jets accompanied by a pair $e^+e^-$: QCD processes, which
depend exclusively on the strong interaction; electroweak processes,
which contains the Drell-Yan production of quark or lepton pairs and the
single and pair productions of electroweak gauge bosons; and the
production of top quark pairs.

The signals for leptoquark production and their associated backgrounds
exhibit different kinematical distributions for the hardest leptons
and jets.  For instance \cite{efl}, the transverse momentum
distributions of the $e_{1(2)}$ and $j_{1(2)}$ for the signal exhibit
a larger fraction of very hard jets and leptons than the backgrounds.
The invariant mass of pairs $e^+ e^-$ is usually quite large for
single and pair productions of leptoquarks, while this spectrum for
the background is concentrated at small invariant masses and around
the mass of the $Z$. Furthermore, the signal events present a clear
peak in the invariant mass ($M_{ik}$) distribution of $e_i$--$j_k$
pairs, as expected.

Taking into account the features of the signal and backgrounds above
described, we imposed the following set of cuts:

\begin{itemize}
  
\item [(C1)] The leading jets and $e^\pm$ should be in the
  pseudorapidity interval $|\eta| < 3$.
        
\item [(C2)] The leading leptons ($e_1$ and $e_2$) should have $p_T >
  200$ GeV.
      
\item [(C3)] We reject events where the invariant mass of the pair
  $e^+e^-$ ($M_{e_1 e_2}$) is smaller than 190 GeV.

\item [(C4)] In order to further reduce the $t\bar{t}$ background
 we required that {\em all} the invariant masses $M_{e_i j_k}$ are
 larger than 200 GeV.

\end{itemize}

We demonstrate in Ref.\ \cite{efl} that the cuts C1---C4 effectively
suppress all the backgrounds to leptoquark production at the LHC,
consequently, the leptoquark searches are background free.  Therefore,
the LHC will be able to exclude at 95\% CL the regions of parameter
space where the number of expected signal events is larger than 3 for
a given integrated luminosity.

The limits on leptoquarks coming from the leptoquark pair searches
depend exclusively on their branching ratio into a charged lepton and
jet ($B_e$) since they are produced by strong interaction processes
that are the same for all leptoquark species. We show in Table~
\ref{t:lim:pair} the 95\% CL limits on the leptoquark masses that can
be obtained from their pair production at the LHC for two different
integrated luminosities. As we can see, this search will be able to
exclude leptoquarks with masses up to 1.5 (1.7) TeV for $B_e = 0.5$
(1) and an integrated luminosity of 100 fb$^{-1}$.

%%%

\begin{table}
\begin{center}
\begin{tabular} {||c|c|c||}
\hline \hline
leptoquark & ${\cal L}=10$ fb$^{-1}$ &${\cal L}=100$ fb$^{-1}$  
\\
\hline \hline
$S_{1L} $ and $S^0_3$   & 1.1 & 1.5 \\
$S_{1R} $,   $\tilde{S}_{1R}$, $R_{2L}^1$,$R_{2R}^2$, and
$\tilde{R}_2^1$  & 1.3 & 1.7 
\\
\hline \hline
\end{tabular}
\vskip 0.75cm
\caption{
  95\% CL limits on the leptoquark masses in TeV that can be obtained
  from the search for leptoquark pairs for two integrated
  luminosities.  }
\label{t:lim:pair}
\end{center}
\end{table}

%%%

We also analyzed the search for leptoquarks through the existence of
an excess of events presenting a $e$-jet invariant mass in the range
$| M_{lq} \pm \Delta M|$ after we imposed the cuts C1--C4 . These
events originate from the single production of leptoquarks, as well as
their pair production, consequently having a larger cross section than
the pair or single production alone. However, this search is model
dependent since the single production involves the couplings of the
leptoquark to fermions.

\begin{figure}
%%%\begin{tabular}{p{0.33\linewidth}p{0.33\linewidth}p{0.33\linewidth}}
\begin{center}
  \mbox{\epsfig{file=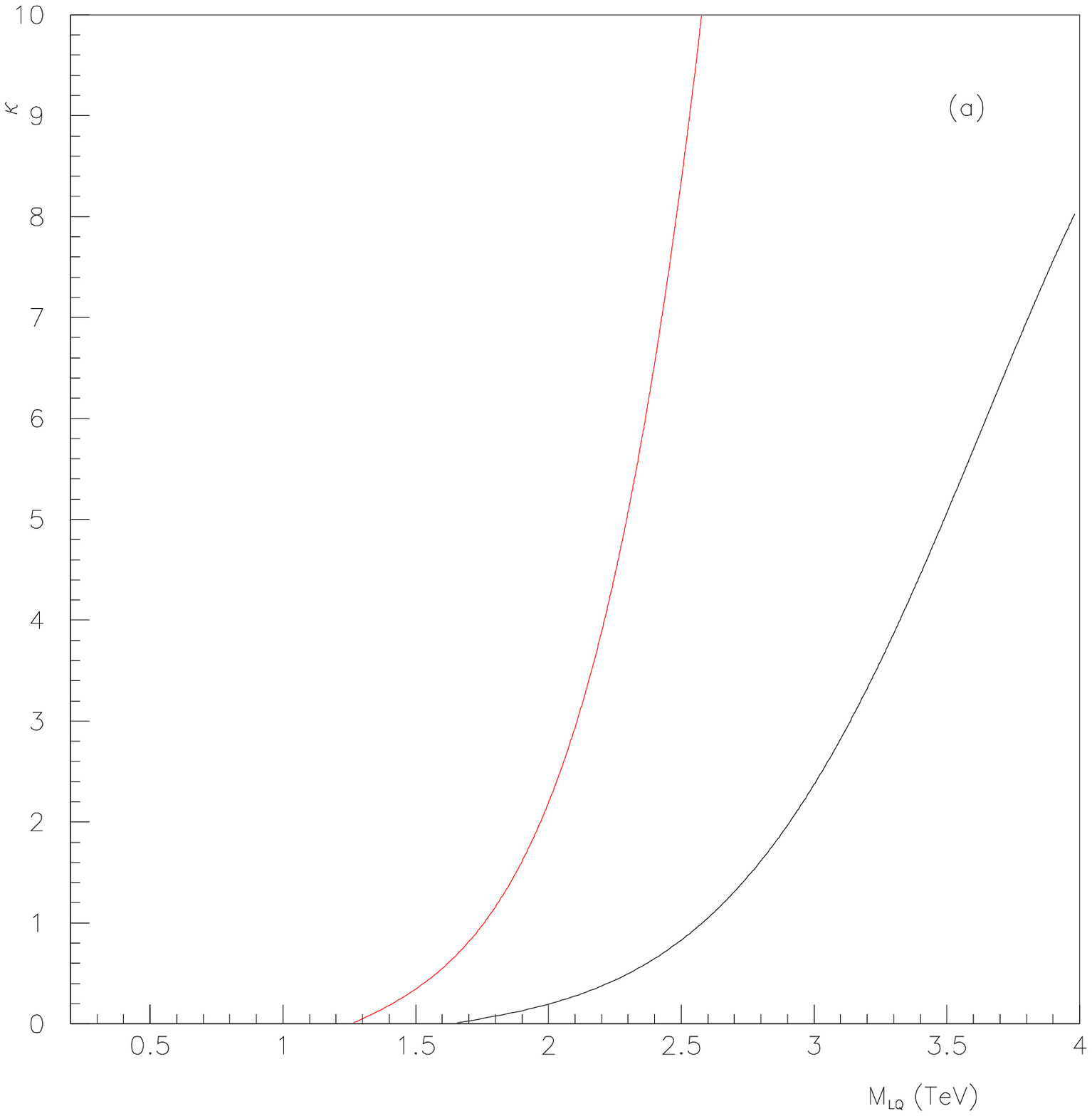,width=0.4\linewidth}}  %%%0.95
\end{center}
%%% &
\begin{center}
  \mbox{\epsfig{file=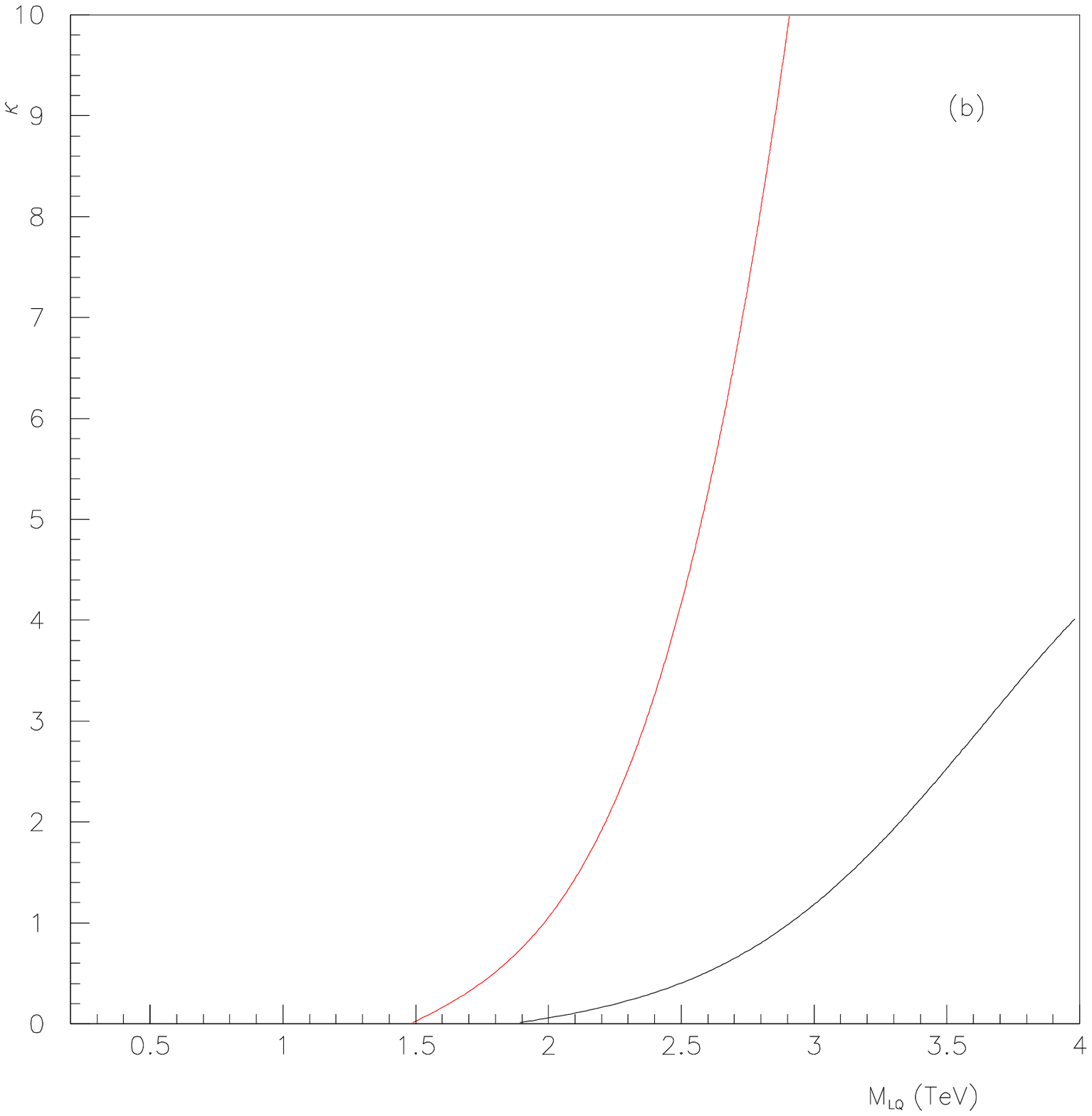,width=0.4\linewidth}}
\end{center}
%%% &
\begin{center}
  \mbox{\epsfig{file=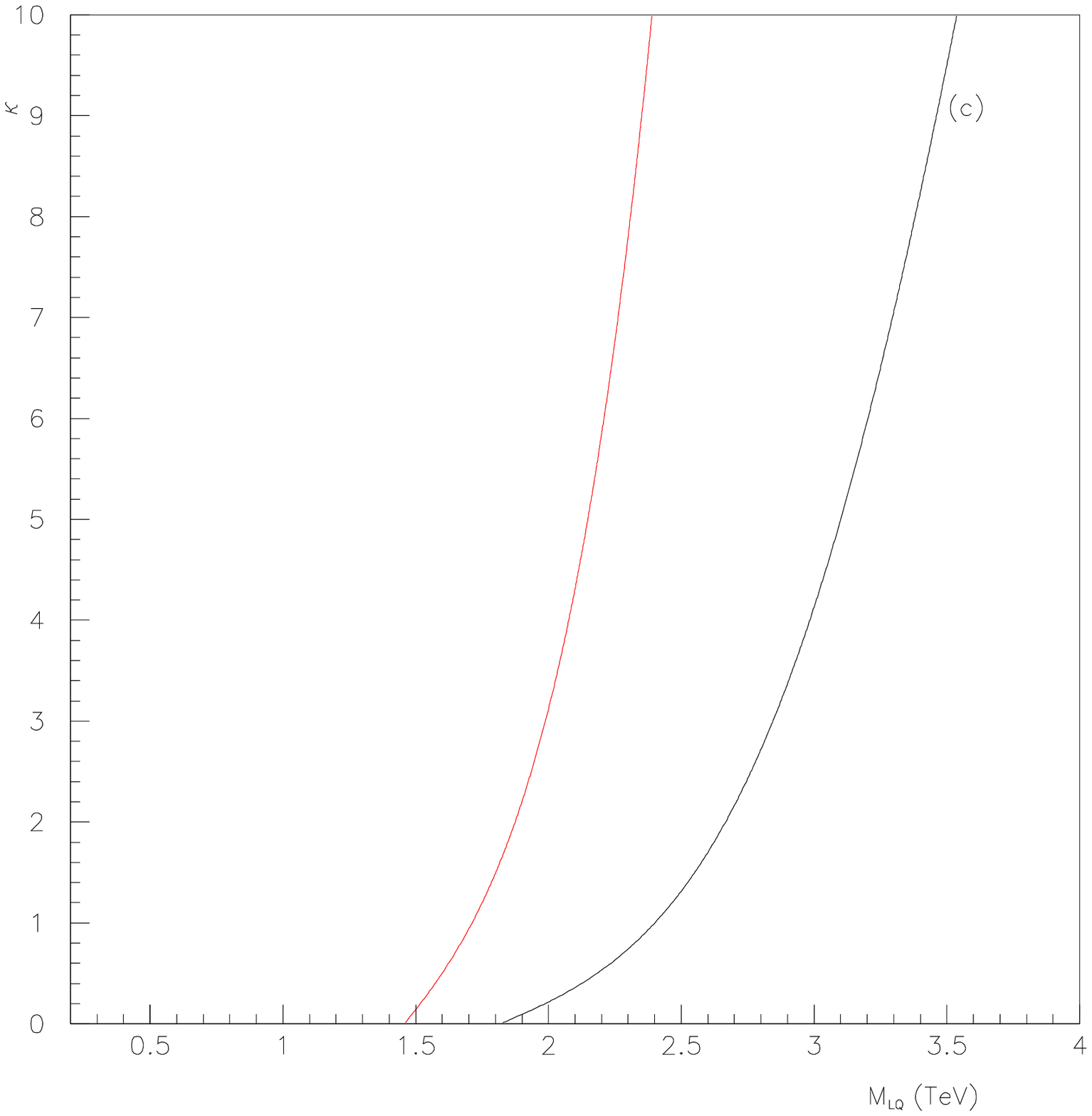,width=0.4\linewidth}}
\end{center}
%%% \\
%%% \end{tabular}
\caption{
  95\% excluded regions in the plane $\kappa$--$M_{lq}$ from the
  single leptoquark analysis for an integrated luminosity of 10/100
  fb$^{-1}$ (solid/dotted line) and the leptoquarks: (a) $S_{1L}$ and
  $S_3^0$; (b) $S_{1R}$, $R^1_{2L}$, and $R^1_{2R}$; (c)$S^+_3$,
  $R^2_{2R}$, $\tilde{R}^1_2$, and $\tilde{S}_{1R}$.  }
\label{kap_mlq}
\end{figure}

Fig.\ \ref{kap_mlq} contains the 95\% CL excluded regions in the plane
$\kappa$--$M_{lq}$ ($\kappa \alpha_{{em}}\equiv {\lambda^2}/{4\pi}$)
from the single leptoquark search.  As expected, the excluded region
is independent of $\kappa$ for masses up to the reach of the
leptoquark pair searches. At higher masses the signal is dominated by
the single production and consequently the bounds on leptoquarks that
couple to $d$ quarks ($S^+_3$, $R^2_{2R}$, $\tilde{R}^1_2$, and
$\tilde{S}_{1R}$) are the weakest ones for a fixed value of
$\kappa$. Since the leptoquarks $S_{1R}$, $R^1_{2L}$, and $R^1_{2R}$
couple to $u$ quarks and have $B_e=1$ they are the ones that possesses
the most stringent limits.  In fact, for leptoquark Yukawa couplings
of the electromagnetic strength ($\kappa=1$) and an integrated
luminosity of 100 fb$^{-1}$, the LHC can exclude $S_{1L}$ and $S_3^0$
leptoquarks with masses smaller than 2.6 TeV; while $S^+_3$,
$R^2_{2R}$, $\tilde{R}^1_2$, and $\tilde{S}_{1R}$ leptoquarks with
masses smaller than 2.4 TeV can be ruled out; and $S_{1R}$,
$R^1_{2L}$, and $R^1_{2R}$ leptoquarks can be excluded up to masses of
2.9 TeV.

Our conclusions are that the LHC will be able to discover leptoquarks
with masses smaller than 1.5--1.7 TeV irrespective of their Yukawa
couplings through their pair production for an integrated luminosity
of 100 fb$^{-1}$.  Furthermore, the single leptoquark search can
extend the reach of the LHC, allowing the discovery of leptoquarks
with masses up to 2--3 TeV depending on their couplings to fermions.

%******************************************************************************

\section{Acknowledgements}

This work was supported by Conselho Nacional de Desenvolvimento
Cient\'{\i}fico e Tecnol\'ogico (CNPq) and by Funda\c{c}\~ao de Amparo
\`a Pesquisa do Estado de S\~ao Paulo (FAPESP).

%******************************************************************************

%******************************************************************************
%Figures
%%%%%%%%%

%%

\end{document}